\documentclass[preprint,review,12pt]{elsarticle}
\usepackage{amssymb,amsmath}

\usepackage{graphicx}
\usepackage{float}

\journal{Physica D}

\begin{document}

\begin{frontmatter}


\title{Quasibreathers in MMT model}

\author[nsu,ws,lpi]{Pushkarev A.\corref{cor1}}
\ead{dr.push@gmail.com}

\author[nsu,ws,lpi,dma]{Zakharov V.E.}
\ead{zakharov@math.arizona.edu}

\cortext[cor1]{Corresponding author}

\address[nsu]{Novosibirsk State University, Novosibirsk, 630090, Russia}
\address[ws]{Waves and Solitons LLC, 1719 W. Marlette Ave., Phoenix, AZ 85015, USA}
\address[lpi]{Lebedev Physical Institute RAS, Leninsky 53, Moscow 119991, Russia}
\address[dma]{Department of Mathematics, University of Arizona, Tucson, AZ 85721, USA}

\begin{abstract}
We report numerical detection of new type of localized structures in the frame of Majda-McLaughlin-Tabak ($MMT$) model  adjusted for description of essentially nonlinear gravity waves on the surface of ideal deep water. These structures -- quasibreathers, or oscillating quasisolitons -- can be treated as groups of
freak waves closely resembling experimentally observed "Three Sisters" wave packs on the ocean surface. The $MMT$ model has quasisolitonic solutions. Unlike  $NLSE$ solitons, $MMT$ quasisolitons are permanently backward radiating energy, but nevertheless do exist during thousands of carrier wave periods. Quasisolitons of small amplitude are regular and stable, but large-amplitude ones demonstrate oscillations of amplitude and spectral shape. This effect can be explained by periodic formation of weak collapses, carrying out negligibly small amount of energy. We call oscillating quasisolitons "quasibreathers".
\end{abstract}

\begin{keyword}
Nonlinear Schr\"{o}dinger Equation \sep solitons \sep freak waves \sep singularities \sep breathers
\end{keyword}

\end{frontmatter}

\section{Introduction}
\begin{figure}[ht]
\includegraphics[scale=1.]{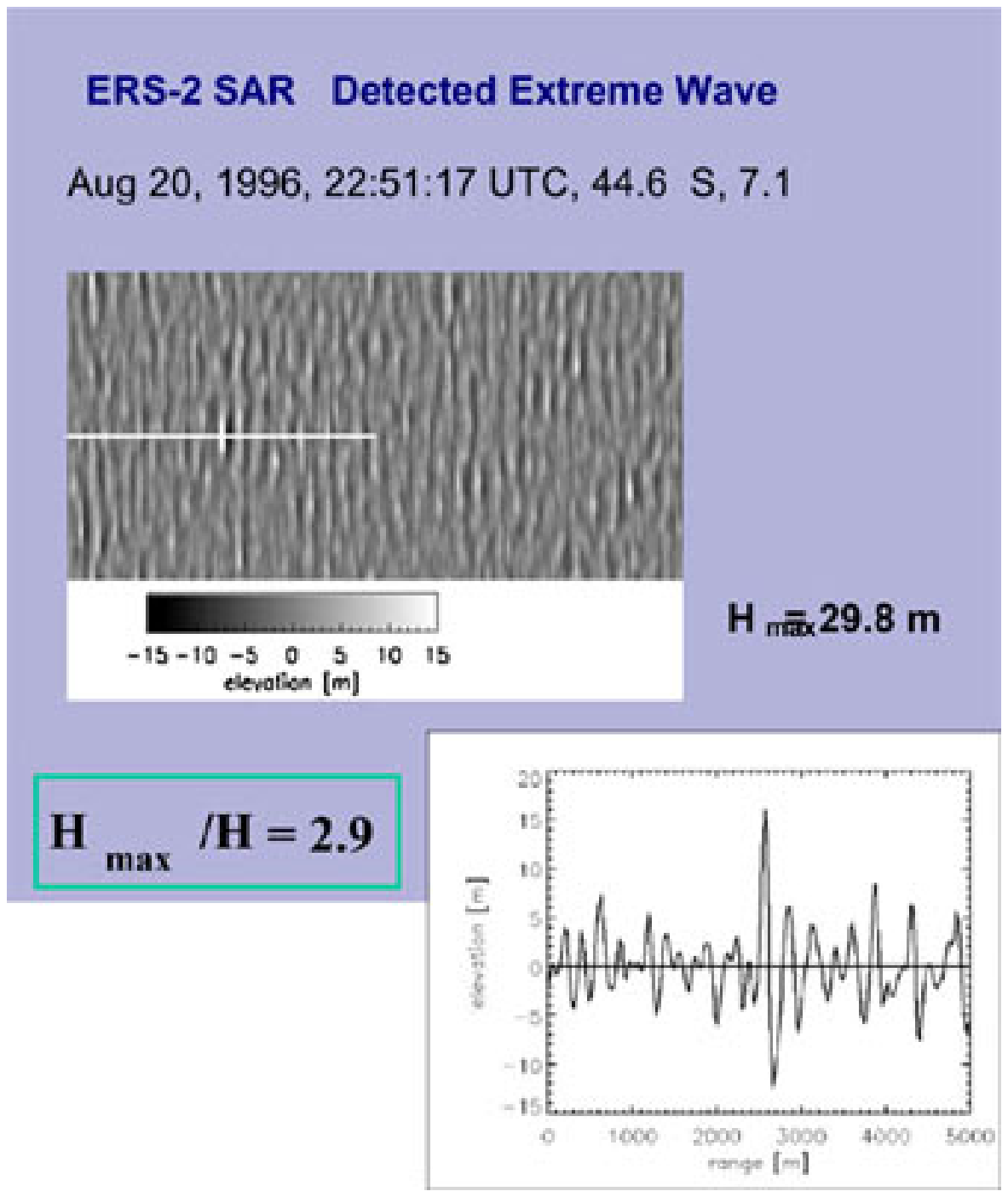} 
\caption{Giant wave detected during a global census using three weeks of raw ERS-2 SAR imagette data, carried out by the German Aerospace Centre (DLR). This SAR data set was inverted to individual wave heights and investigated for individual wave height and steepness. The wave shown here has a height of 29.8 m. 
Adopted from $ http://www.esa.int/esaCP/SEMOKQL26WD \_ index \_ 1.html\sharp subhead4$
}\label{SAT}
\end{figure}

\begin{figure}[ht]
\includegraphics[scale=0.35]{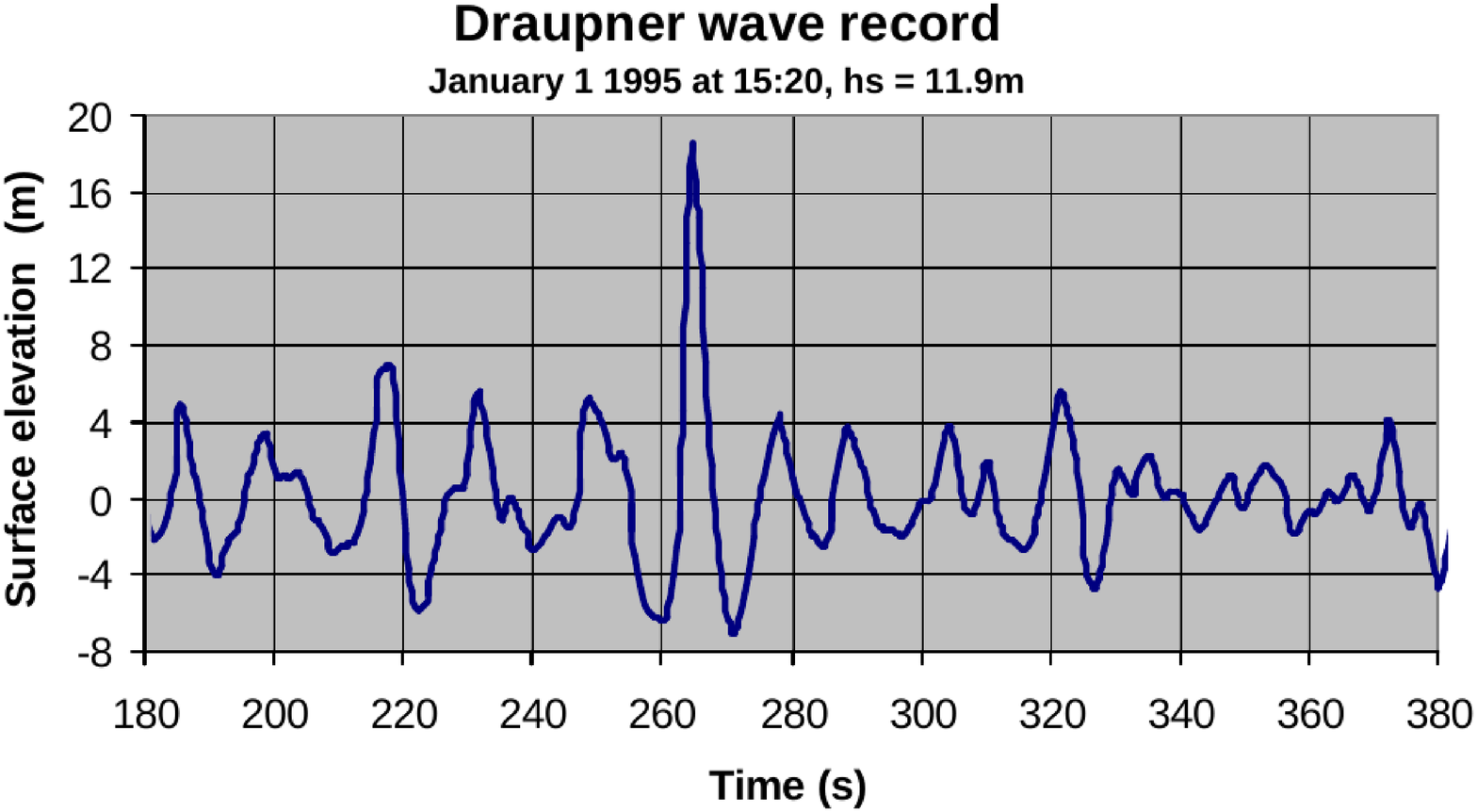} 
\caption{Freak wave event detected from the Draupner oil platform on Jan.1, 1995. Adopted from http://www.math.uio.no/~karstent/seminarV05/Haver2004.pdf}\label{Draupner}
\end{figure}

Development of analytic theory of freak (or rogue) waves is one of the most interesting problems of hydrodynamics. In spite of recent progress in this area \cite{PK} many important questions are not answered yet. Apparently, freak waves are the structures well localized in space, see Fig.\ref{SAT}. But behavior of freak waves in time in co-moving coordinate frames is not still explored. From the experimental viewpoint this is a hard question. It cannot be answered by a resting observer, for whom the freak wave is just a single event localized in time, see Fig.\ref{Draupner}. From the other hand, satellites move too fast to record the full "live story" of a freak wave.

The standard model for description of freak waves in deep water is Nonlinear Schr\"{o}dinger Equation ($NLSE$). This equation has a plethora of exact solutions which often are associated with the freak waves on deep water. Some of these solutions are presented in \cite{AA}, more recent developments can be found in articles \cite{TW}-\cite{A2}. These solutions, however, presume existence of background monochromatic wave (condensate) and are connected to the subject of our paper only indirectly. For this reason, we do not pursue a purpose to present here the detailed description of all solitonic solutions on the condensate background, as well as completed and controversial history of their discovery. In this article we study the solitons on almost zero background.

However, we should mention the remarkable $NLSE$ solution found by Peregrine \cite{P}. This solution in the co-moving coordinate frame is an instanton, describing the single event -- appearance and disappearance of the freak waves group. Today we can speak about two alternative versions of the freak wave theory. The "instantonic" version assumes that the freak wave is a single event, localized in time. The "solitonic" version proposes that the freak wave are described by persistent solitons, probably oscillating in time. So far experimental data are too scarce to make a conclusion in favor of one of these theories. 

One should remember that $NLSE$ is derived in the assumption that the wave train size, containing the freak waves, is much larger than characteristic wave length. Most of collected experimental data, however, show that in the real ocean this condition is not satisfied (see Fig.\ref{SAT}, \ref{Draupner}) and that the $NLSE$ is hardly applicable. 

A level of nonlinearity of  quasi-monochromatic wave group is measured by the characteristic steepness $ \mu \simeq ka $ ($k$ is the wavenumber and $a$ is the amplitude). Our numerical experiments  \cite{PDZ} show that $NLSE$ is applicable if $\mu \lessapprox 0.07$. According to our calculations, $NLSE$ is not applicable if $\mu \simeq 0.1$. Recent numerical experiments \cite{S} show that this limit might be extended to $\mu\simeq 0.15$. However, for freak waves in the real sea $\mu \simeq 0.3 \div 0.5$ (see Appendix II). The $NLSE$ is absolutely not applicable for description of that steep freak waves.

It is also known from observations \cite{PK} that a typical configuration of a freak wave group consists of three sequent waves -- "Three Sisters". This group is too short to be described by $NLSE$.

What are the alternatives to $NLSE$ model? The most consistent approach is the use of the exact Euler equations for description of the potential flow of the ideal fluid with free surface. Some advances in this direction are already achieved \cite{PDZ}-\cite{DZ}. However, the study of more simple and less accurate models also could be very useful. In this article we present our result on numerical solution of well-known $MMT$ \cite{MMT} equation with the special choice of parameter $\alpha=1/2, \beta=3, \lambda=+1$, making this model well adjusted for description of surface gravity waves.

Our results mostly support the "solitonic" theory of freak waves. We started with initial data, corresponding to $NLSE$ solitons and discovered formation of persistent quasisolitons existing for more than two thousands of wave periods. These quasisolitons slowly radiate energy in backward direction. As was shown recently \cite{RNZ} in the "model case" ($\alpha=1/2, \beta=0, \lambda=1$), this effect plays the key role in formation of the wave turbulent spectrum, but in our case its influence is negligibly small. 

However, we discovered completely new effect. While quasisolitons of small steepness ($\mu \lessapprox 0.1$) behave similar to $NLSE$ solitons on zero background, the quasisolitons of higher steepness demonstrate almost periodic oscillations of amplitude and spectral shape, periodically forming power-like tails in spectra. This effect can be explained by modulational instability inside the quasisoliton. Development of this instability leads to formation of "weak" one-dimensional collapses, which deform the spectrum, but absorb negligibly small amount of energy. Thereafter we call oscillating quasisolitons "quasibreathers".

\section{Basic model}

The Majda-McLaughlin-Tabak ($MMT$) equation (see \cite{MMT}, \cite{ZDP} and \cite{ZGPD})
\begin{eqnarray}
\label{MMT}
i \frac{\partial \psi}{\partial t} &=& \left| \frac{\partial }{\partial x} \right|^\alpha \psi + \lambda \left| \frac{\partial}{\partial x} \right|^{\beta/4} \left( \left|\left| \frac{\partial }{\partial x} \right|^{\beta/4} \psi \right|^2 \left| \frac{\partial }{\partial x} \right|^{\beta/4} \psi \right), \\
\lambda &=& \pm 1 ,\,\,\,-\infty<x<\infty, \,\,\,\, 0<t<\infty \nonumber
\end{eqnarray}
where $\psi(x,t)$ is the complex function and the fractional derivative is defined by
\begin{eqnarray}
\label{FT}
\left| \frac{\partial }{\partial x} \right|^\alpha \psi = \int |k|^\alpha \psi_k e^{ikx} dk
\end{eqnarray}
has been attracting lately fare attention of nonlinear wave scientists. The reason is $MMT$ equation incorporates several already known important cases, and also can be used as a ``test-bed'' for verification of the concepts like weak-turbulent waves spectra, localized structures and their co-existence \cite{ZDP}, \cite{ZGPD}. 
For $\alpha=0$ and $\beta=0, 2$ Eq.\ (\ref{MMT}) is completely integrable. If $\alpha=2$ and $\beta=0$, it is the classical $NLSE$ for focusing ($\lambda=-1$) and defocusing ($\lambda=+1$) cases: 
\begin{eqnarray}
\label{NLS}
i \frac{\partial \psi}{\partial t} = -\frac{\partial^2 \psi}{\partial x^2} + \lambda |\psi|^2 \psi
\end{eqnarray}
If $\alpha=2$ and $\beta=2$, transformation $\phi=|\frac{\partial}{\partial x}|^{\frac{1}{2}}\psi$ turns Eq.(\ref{MMT}) into the derivative $NLSE$ \cite{Kundu}:
\begin{eqnarray}
\label{NLES}
i \frac{\partial \phi}{\partial t} = -\frac{\partial^2 \phi}{\partial x^2} + \lambda \frac{\partial}{\partial x} |\phi|^2 \phi \nonumber
\end{eqnarray}
Through Fourier transform 
\begin{equation}
\psi_k = \frac{1}{2\pi}\int\psi(x) e^{-ikx}dx \nonumber
\end{equation}
Eq.(\ref{MMT}) can be rewritten in the form
\begin{eqnarray}
\label{MMT_FFT} 
i\frac{\partial \psi_k}{\partial t} = |k|^\alpha \psi_k+\int T_{k k_1 k_2 k_3} \psi_{k_1}^\star \psi_{k_2}\psi_{k_3} \delta_{k+k_1+k_2+k_3} dk_1 dk_2 dk_3
\end{eqnarray}
where 
\begin{eqnarray}
\label{MMT_ME}
T_{k k_1 k_2 k_3} = \lambda |k|^{\beta/4} |k_1|^{\beta/4} |k_2|^{\beta/4} |k_3|^{\beta/4}
\end{eqnarray}
Suppose that in Eq.\ (\ref{MMT_FFT}) $T_{k k_1 k_2 k_3}$ is a generic function satisfying the symmetry conditions
\begin{eqnarray}
\label{cond}
T_{k k_1, k_2 k_3} = T_{k_1 k, k_2 k_3} = T_{k k_1, k_3 k_2} = T_{k_2 k_3, k k_1} 
\end{eqnarray}
For matrix coefficient (\ref{MMT_ME}) conditions (\ref{cond}) are satisfied and Eq.(\ref{MMT}) is a Hamiltonian system
\begin{eqnarray}
\label{H}
i\frac{\partial \psi_k}{\partial t} &=& \frac{\delta H}{\delta \psi_k^*}, \nonumber \\ 
H&=&\int |k|^\alpha |\psi_k|^2 dk + \frac{1}{2} \int T_{k k_1 k_2 k_3} \psi_k^\star \psi_{k_1}\psi_{k_2} \psi_{k_3} \delta_{k+k_1-k_2-k_3} dk dk_1 dk_2 dk_3 \nonumber
\end{eqnarray}
Obviously, the Hamiltonian $H$ is a constant of motion. Other motion constants are wave action
\begin{eqnarray}
\label{N}
N = \int \left| \psi_k \right|^2 dk \nonumber
\end{eqnarray}
and wave momentum
\begin{eqnarray}
\label{M}
P = \frac{i}{2} \int\left( \psi \frac{\partial \psi^\star}{\partial x}  - \frac{\partial \psi}{\partial x} \psi^\star \right) dx \nonumber
\end{eqnarray}
Another model of type (\ref{MMT_FFT}), describing surface waves on deep water, is so-called ``Zakharov equation'' \cite{ZE}. This equation is not heuristic  like $MMT$, it was systematically derived from Euler equations and therefore is supposed to be more accurate in corresponding context. In this equation $T_{\epsilon k,\epsilon k_1,\epsilon k_2,\epsilon k_3} = \epsilon^3 T_{k k_1 k_2 k_3}$ is cumbersome homogeneous function of the third order.

One should note that if Eq.\ (\ref{MMT}) is applied for description of  gravity waves, the surface shape can be reconstructed by the formula (see {\bf Appendix I})
\begin{eqnarray}
\label{SurfaceElevations}
\eta(x,t) = \frac{1}{\sqrt{2}} \int e^{ikx} |k|^{1/4} (\psi_k+\psi_k^*)dk
\end{eqnarray} 

\section{Solitons and quasisolitons}
\begin{figure}[ht]
\includegraphics[scale=0.15]{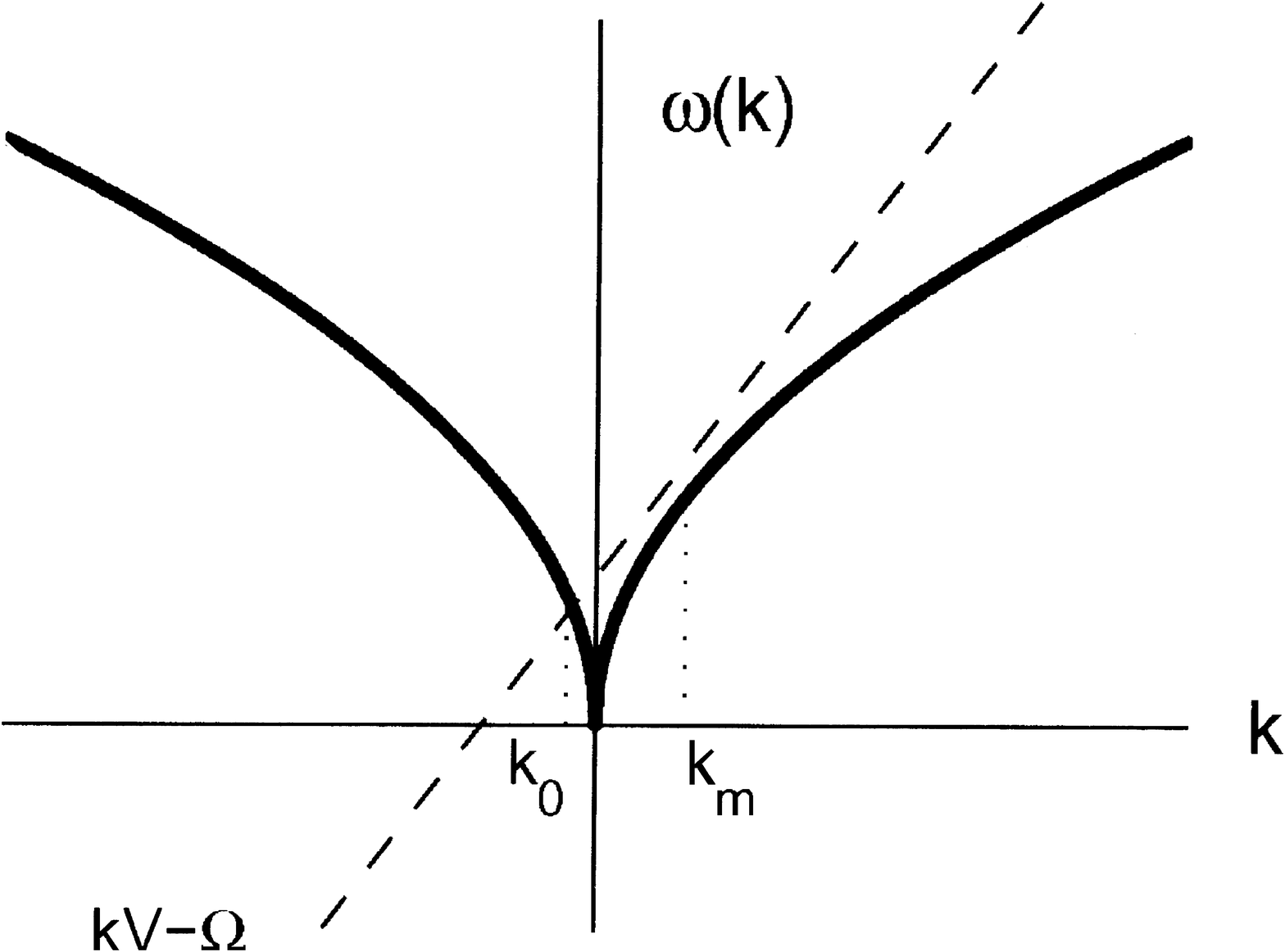} 
\caption{Example of the situation when defocusing quasisolitons are possible. The dispersion relation is $\omega=|k|^\alpha$ for $\alpha<1$, $\Omega$ is negative and $V$ is positive. The straight line always crosses the dispersion relation $\omega=\omega(k)$ and, therefore, the denominator $\Omega-kV+\omega(k)$ in Eq.\ (\ref{Sol}) has zero. Quasisoliton takes place only in the defocusing case $\lambda=+1$.} \label{DisRel}
\end{figure}

Let us look for a solution of Eq.(\ref{MMT_FFT}) in a form
\begin{eqnarray}
\label{GenSol}
\psi_k(t) = e^{i(\Omega-kV)t} \phi_k
\end{eqnarray}
where $\Omega$ and $V$ are the constants. The function $\phi_k$ should satisfy the nonlinear integral equation
\begin{eqnarray} 
\label{Sol}
\phi_k= \lambda \frac{\int T_{1234} \phi_1^\star \phi_2 \phi_3 \delta(k+k_1-k_2-k_3)dk_1dk_2dk_3}{-\Omega+kV-|k|^\alpha}
\end{eqnarray}
This equation has solutions if $\Omega$ and $V$ can be chosen such that the denominator in Eq. (\ref{Sol}) cannot be zero for real $k$. This might happen only if $\alpha > 1$.
Let's suppose now $\alpha<1$. One can see that in this case the denominator in Eq.\ (\ref{Sol}) always has zero, which is clear from Fig.\ref{DisRel}. Let $\Omega<0$, $V>0$.
\noindent
Thus, any solution of type (\ref{GenSol}) has singularity at negative $k$. It means that strict soliton solution of type (\ref{GenSol}) does not exist. However, one can construct approximate solutions, such that $\phi_k$ in (\ref{GenSol}) is slow function of time. These approximate moving solutions, radiating energy in the backward direction are called quasisolitons after paper \cite{ZK}.

As it was recently shown in \cite{ZDP}, \cite{ZGPD}, quasisolitons play the central role in wave turbulence in the frame of $MMT$ model if $\alpha=\frac{1}{2}$, $\beta=0$ and $\lambda=+1$. It was shown that in this case the backward radiation plays the central role in dynamics of quasisolitons. But we study only the case $\alpha=\frac{1}{2}$, $\beta=3$ and $\lambda=+1$. In this case , which intentionally models the gravity waves on deep water, the backward radiation is not that strong, due to essential nonlinearity suppression in the area of small wave numbers. Nevertheless, we  definitely detect this phenomenon in our numerical experiments.

Consider the structure of the denominator in Eq.(\ref{Sol}). One can expect existence of the quasisoliton in the case when the straight line $\omega = k V-\Omega$ is tangential to the curve $\omega = k^\alpha$. The conditions of equal derivatives and existence of the common point of these two curves at $k=k_m$ are:
\begin{eqnarray}
\label{Tangent}
V &=& \alpha k_m^{\alpha-1} \\
\label{CommonPoint}
\Omega &=& (\alpha-1) k_m^\alpha
\end{eqnarray}
We are now returning back to non-stationary Eq.(\ref{MMT_FFT}) and make the change of variables $k = k_m +\kappa$, $\kappa<<k$. Dispersion relation expansion into Taylor series
\begin{eqnarray}
(k_m+\kappa)^\alpha= k_m^\alpha+\alpha k_m^{\alpha-1}\kappa+\frac{1}{2}\alpha(\alpha-1)k_m^{\alpha-2}\kappa^2 \nonumber
\end{eqnarray}
and change of variables 
\begin{eqnarray}
\psi_k (t) = e^{-i (k_m^\alpha + \alpha k_m^{\alpha-1}\kappa)t} \phi_{\kappa}(t) 
\end{eqnarray}
gives
\begin{eqnarray}
&&i\frac{\partial \phi_\kappa}{\partial t}= \frac{1}{2}\alpha(\alpha-1) k_m^{\alpha-2} \kappa^2 \phi_\kappa +\\
&+& k_m^\beta \int \phi_{\kappa_1}^\star \phi_{\kappa_2}\phi_{\kappa_3} \delta(\kappa+\kappa_1-\kappa_2-\kappa_3) d\kappa_1 d\kappa_2 d\kappa_3 = 0
\end{eqnarray}
Another change of variables
\begin{eqnarray}
\phi_\kappa = e^{i \Delta  t} \chi_\kappa, \,\,\,\,\,\,\, \Delta = \frac{1}{2}\alpha(\alpha-1) k_m^{\alpha-2}q^2 \nonumber
\end{eqnarray}
gives
\begin{eqnarray}
&&i \frac{\partial \chi_\kappa}{\partial t} = \frac{1}{2}\alpha(\alpha-1) k_m^{\alpha-2}(q^2 + \kappa^2) \chi_\kappa +\\
&+& i k_m^\beta \int \chi_{\kappa_1}^\star \chi_{\kappa_2}\chi_{\kappa_3} \delta(\kappa+\kappa_1-\kappa_2-\kappa_3) d\kappa_1 d\kappa_2 d\kappa_3 \nonumber
\end{eqnarray}
Applying inverse Fourier transform $\chi(x,t) = \int \chi_\kappa(t) e^{i \kappa x} d \kappa$ to the last equation, we get $NLSE$ in real space:
\begin{eqnarray}
\label{NLS-chi-Real}
i \frac{\partial \chi}{\partial t} + \frac{1}{2}\alpha (1-\alpha) k_m^{\alpha-2}(q^2 \chi - \frac{\partial^2 \chi}{\partial x^2}) - k_m^\beta |\chi|^2 \chi = 0
\end{eqnarray}
Eq.(\ref{NLS-chi-Real}) has partial stationary solution
\begin{eqnarray}
\label{NLS-stationary-soliton}
\chi(x) = \sqrt{\frac{\alpha (\alpha-1)}{k_m^{\beta-\alpha+2}}} \frac{q}{\cosh{ q x }}
\end{eqnarray}
which produces approximate quasisoliton solution of the Eq.(\ref{MMT}) with $\lambda = 1$:
\begin{eqnarray}
\label{MMT-quasi-soliton}
\psi(x,t) &=& \chi(x-vt) e^{i (\Omega + \Delta) t} e^{i k_m (x-vt)} \\
\Omega &=& - (1-\alpha) k_m^\alpha \nonumber \\ 
\Delta &=& - \frac{1}{2} \alpha (1-\alpha) k_m^{\alpha-2} q^2 \nonumber \\
V &=& \alpha k_m^{\alpha-1} \nonumber
\end{eqnarray}



The characteristic wave-number $k_0=-c k_m$ of backward radiation associated with the quasisoliton (see Fig.\ref{DisRel}) can be found from the equation
\begin{eqnarray}
\label{Zeroes}
k_0 V - \Omega = |k_0|^\alpha
\end{eqnarray}
together with Eq.(\ref{Tangent})-(\ref{CommonPoint}). For $\alpha=1/2$
\begin{eqnarray}
\label{C}
c =  3-\sqrt{8} \simeq 0.172
\end{eqnarray}
Therefore, due to the smallness of the ratio $\frac{T(k_0,k_0,k_m,k_m)}{T(k_m,k_m,k_m,k_m)} \simeq c^3=5\cdot 10^{-3}$, the backward radiation process in framework of the $MMT$ model for $\beta=3$ is suppressed with respect to the case $\beta=0$.

To obtain the surface shape we replace in Eq.(\ref{SurfaceElevations}) $k^{1/4}$ with $k_m^{1/4}$ and get 
\begin{eqnarray}
\eta = \frac{q}{k_m} \frac{1}{cosh{q (x-vt)}} \cos(\omega t -k_m x) \nonumber
\end{eqnarray}
Thus $q$ is the standard steepness.

\section{Self-similar collapses}

Eq.(\ref{MMT}) has self-similar solution:
\begin{eqnarray}
\label{PsiSSS}
\psi(x,t) = (t_0-t)^{5/2} F\left(\frac{x}{(t_0-t)^2}\right) 
\end{eqnarray}
For the shape of the surface it gives
\begin{eqnarray}
\label{EtaSSS}
\eta(x,t) = (t_0-t)^2 F\left(\frac{x}{(t_0-t)^2}\right) 
\end{eqnarray}
At $t \rightarrow t_0$ time must vanish from Eq.\ (\ref{MMT}), which means that
\begin{eqnarray}
\eta \rightarrow \alpha^+ x\,\,\,for \, x>0 \nonumber \\
\eta \rightarrow \alpha^- x\,\,\,for \, x<0 \nonumber
\end{eqnarray}
where $\alpha^+>0$ and $\alpha^-<0$ are the constants.
In other words, solution Eq.\ (\ref{PsiSSS}) describes formation of a wedge, in general (if $\alpha^+ \neq \alpha^-$), tilted with respect to vertical line. In $k$-space we get
\begin{eqnarray}
\label{PsiSSS_Kspace}
\psi(k,t) = (t_0-t)^{9/2} F\left(k {(t_0-t)^2}\right) 
\end{eqnarray}
According to (\ref{PsiSSS_Kspace}) $F(\xi) \rightarrow \xi^{-9/4}$ at $\xi \rightarrow 0$. Hence, asymptotically
\begin{eqnarray}
\label{SSS-1}
\psi(k,t) \simeq k^{-9/4} \\
\label{SSS-2}
|\psi(k,t)|^2 \simeq k^{-9/2} 
\end{eqnarray}
Formation of collapses like Eq.\ (\ref{PsiSSS})-(\ref{EtaSSS}) means growth of power-like tails in $k$-space. The spectrum Eq.\ (\ref{SSS-1})-(\ref{SSS-2}) appears only at the moment of collapse $t \rightarrow 0$. The singularity has the form of appearing and vanishing wedge, absorbing some amount of energy. However, time-averaged spectrum can have a slope different from $|\psi_k|^2 \simeq k^{-9/2}$. If the collapse events are rare, the slope must be higher that $k^{-9/2}$.

Let us suppose that the collapse is ``weak'' and that only a very small part if energy is dissipated in an individual event. It means that the collapse is ``almost'' invertible process, symmetric in time with respect to the sign change to $-t$. In other words, the collapsing solution is 
\begin{eqnarray}
\psi(k,t) = |t_0-t|^{9/2}F(k(t_0-t)^2) \nonumber
\end{eqnarray}
Now we can perform the Fourier transform in time and get
\begin{eqnarray}
\psi(k,\omega) = \int_{t_0}^{\infty} |t_0-t|^{9/2} F(k(t_0-t)^2) e^{-i\omega t} dt = e^{i\omega t_0} \frac{1}{k^{11/4}} f\left(\frac{\omega}{k^{1/2}} \right) \nonumber
\end{eqnarray}
The spatial spectrum is given by the integral
\begin{eqnarray}
\label{Spectrum}
I_k = |\psi(k)|^2 \simeq \int|\psi(k,\omega)|^2 d\omega \simeq k^{-5}
\end{eqnarray}
For the surface elevations spectrum we obtain Phillips spectrum
\begin{eqnarray}
|\eta_k|^2\simeq \frac{1}{k^4} \nonumber
\end{eqnarray}
In our numerical experiments we observed the spectra both more steep for $k \rightarrow +\infty$, and more shallow for $k \rightarrow -\infty$ than Eq.\ (\ref{Spectrum}). So far, we have no proper explanation of this fact.

\section{Turbulent quasibreathers in $MMT$ model}

The Eqs.\ (\ref{MMT_FFT})-(\ref{MMT_ME}) have been solved numerically in periodic boundary conditions real space domain $[0,2\pi]$ for deep gravity surface waves case $\alpha=\frac{1}{2}$, $\beta=3$ and $\lambda = 1$. Numerical integration has been performed through iterations of the implicit second order scheme in time and calculation of nonlinear term by Fast Fourier Transform technique. This numerical scheme preserves constants of motion of the approximated equation. 

To avoid high-frequency instabilities, the low-pass filtering has been applied on every time-step through multiplication of the Fourier transform of the wave field by hyper-gaussian function, leaving about $90\%$ of Fourier modes intact, while effectively suppressing the rest of potentially unstable high-frequency modes.  Results were verified against the wave modes number change from $8192$ to  $16384$ and $32768$ for the same Cauchy problem. The calculations were continued typically up to thousands of the initial wave periods without loss of the accuracy.

\begin{figure}[ht]
\includegraphics[scale=0.3]{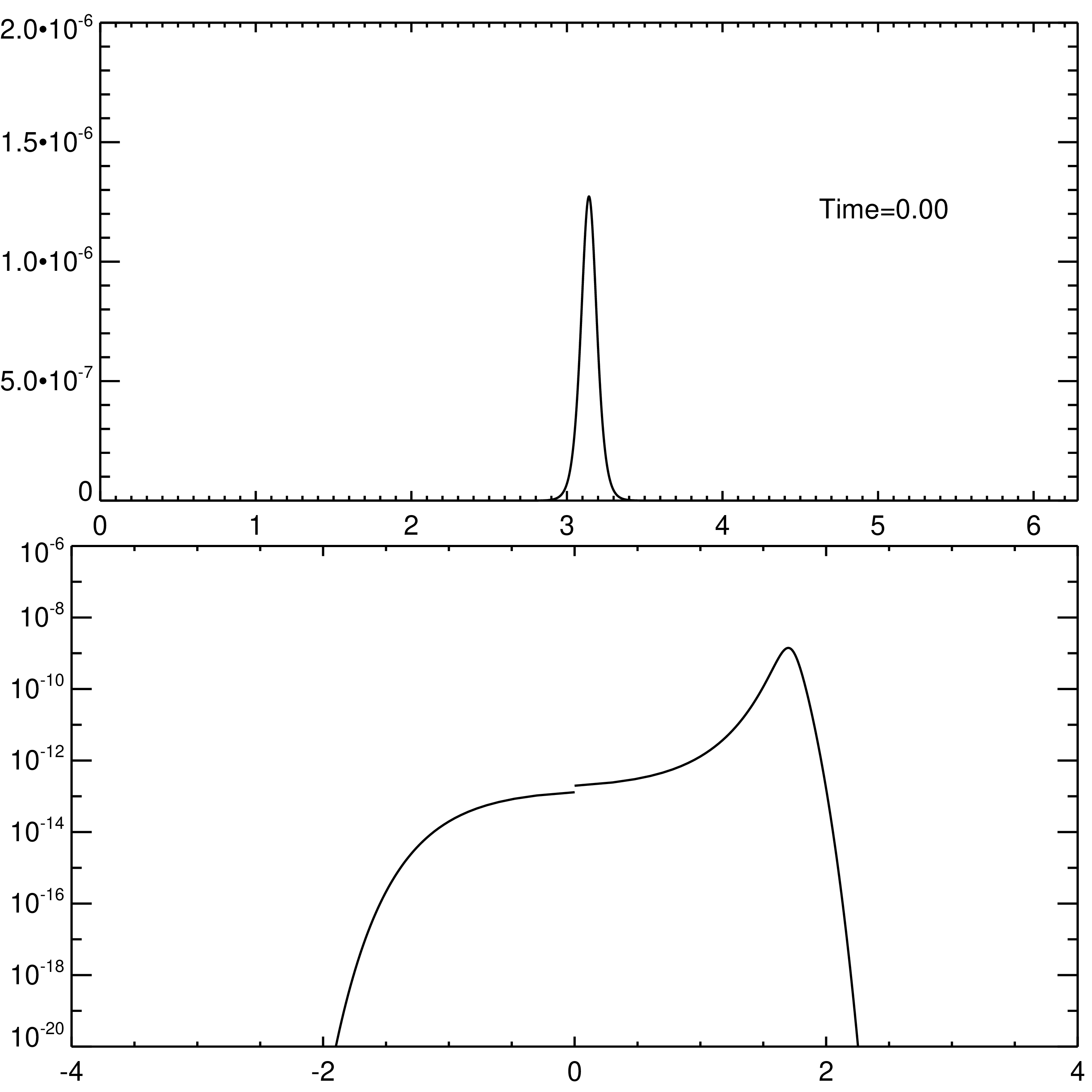} 
\caption{Real and Fourier space distributions of wave field. Top graph: $|\psi(x,t)|^2$ as a function of $x$ for $t=0$. Bottom graph: Fourier spectrum $\log_{10} {|\psi(k,t)|^2}$ as a function of signed logarithm of waves number $\operatorname{sign}(k)\log_{10}{|k|}$ for time $t=0$.} \label{InCond}
\end{figure}

The initial condition was taken in the form of $NLSE$ soliton
\begin{eqnarray}
\label{InCondSoliton}
\psi(x,0) = \frac{q}{2 k_m^{9/4}} \frac{e^{ik_m x}}{\cosh qx}
\end{eqnarray}
for $k_m=50$, see Fig. \ref{InCond}.
It is known \cite{ZGPD} that simulation results essentially depend on the value of the nonlinearity parameter $q/k_m$. For $q/k_m \lesssim 0.1$ the initial condition moves with the constant speed $V$  without any noticeable shape change over characteristic length of at least dozens of simulation domain size $2\pi$.
For $q/k_m > 0.1$, the initial shape Eq.(\ref{InCondSoliton}) starts to change in time and for $q/k_m=0.3$ forms moving wedge-like growing structure with narrowing width. This behavior was interpreted in \cite{ZGPD} as possible collapse of the initial condition over finite time, but further numerical simulation was not continued because of high-wavenumbers instability development in Fourier space, causing blow-up of the numerical scheme.
In current research utilizing more sophisticated numerical approach, it was possible to follow the evolution of the same collapsing initial condition for practically unlimited time. We observed that, in fact, this collapsing initial condition evolves into localized non-stationary solution, periodically recurring to its initial shape. By analogy with cubical $NLSE$, it was interpreted as a breather-like structure. 

\begin{figure}[ht]
\includegraphics[scale=0.5]{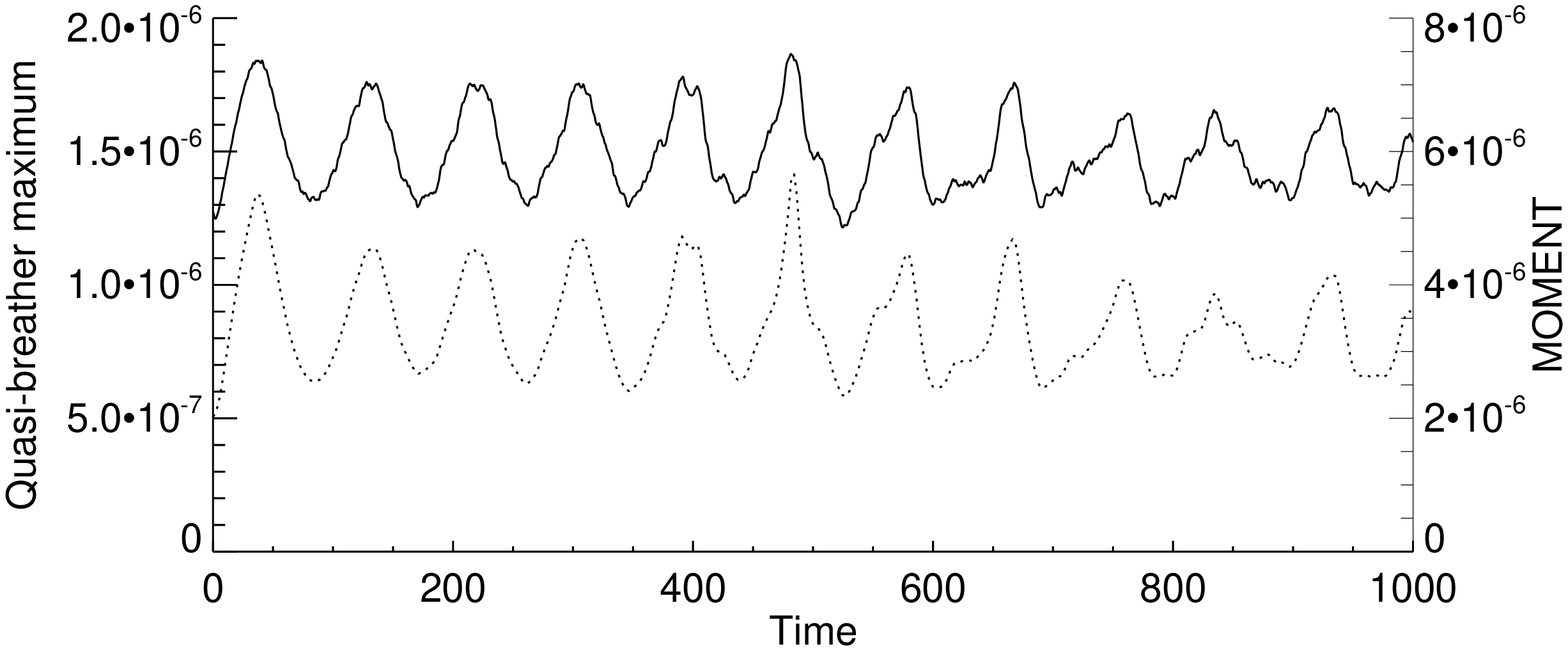} 
\caption{Dependence of the solution maximum $max(|\psi(x,t))^2$,taken over integration domain $[0,2 \pi]$ (solid line, left axis) and the second moment $\int (k-k_0)^2 |\psi_k|^2 dk$ (dotted line, right axis), on time $t$. The average wave-number is defined as $k_0 = \frac{\int k |\psi_k|^2 dk}{\int |\psi_k|^2 dk}$ } \label{OSC_MOM}
\end{figure}

The observed phenomenon is quite interesting: at $q/k_m \sim 0.3$ the initial condition Eq.\ (\ref{InCondSoliton}) evolves into localized object, but with ``inner life''. The shape of this object and the form of its spectra demonstrate irregular, stochastic behavior, which can be interpreted as some ``intrinsic turbulence''.
Time evolution of real space maximum of the solution is presented on Fig.\ \ref{OSC_MOM}. One should note that oscillations are quasi-periodic and their amplitude slowly diminishes in time, at least partially due to destruction of the breather by surrounding noise -- that's the reason why we called this localized state by quasibreather. 
Almost identical picture of oscillations is seen from the second curve on Fig.\ref{OSC_MOM}, which presents the behavior of the second moment as a function of time. Both curves on Fig.\ref{OSC_MOM} clearly indicate the presence of nonlinear oscillating structure in the wave system.

\begin{figure}[ht]
\includegraphics[scale=0.5]{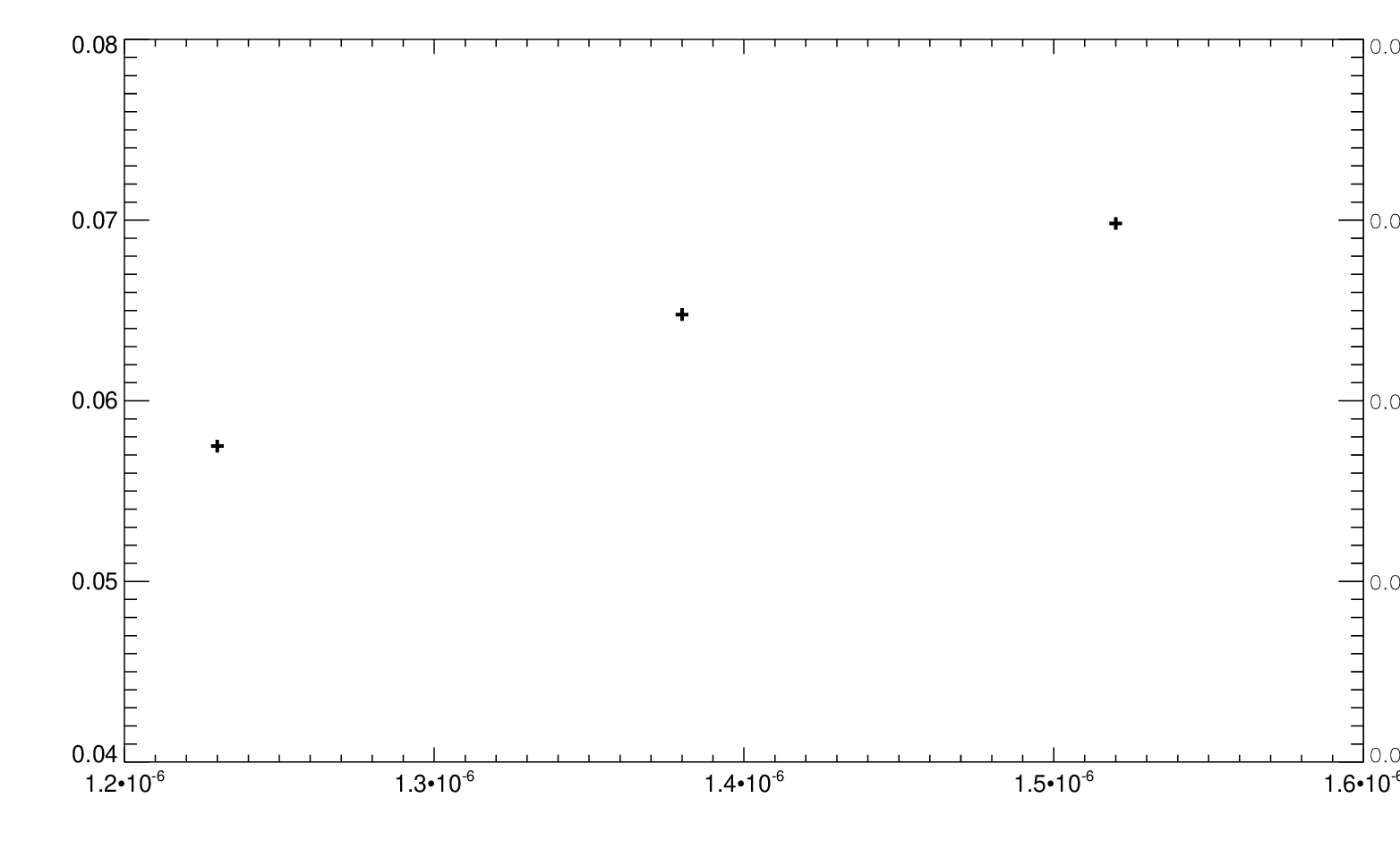} 
\caption{Dependence of quasibreather maximum oscillations frequency on the mean level of these oscillations $<|\psi(x,t)|^2>$.} \label{FreqOsc}
\end{figure}

Fig.\ref{FreqOsc} shows dependence of the frequency of these oscillations on the their mean level. The frequency has a tendency to grow with the growth of the oscillations level. This fact is in a correspondence with frequency dependence on nonlinear frequency shift.

\begin{figure}[ht]
\includegraphics[scale=0.3]{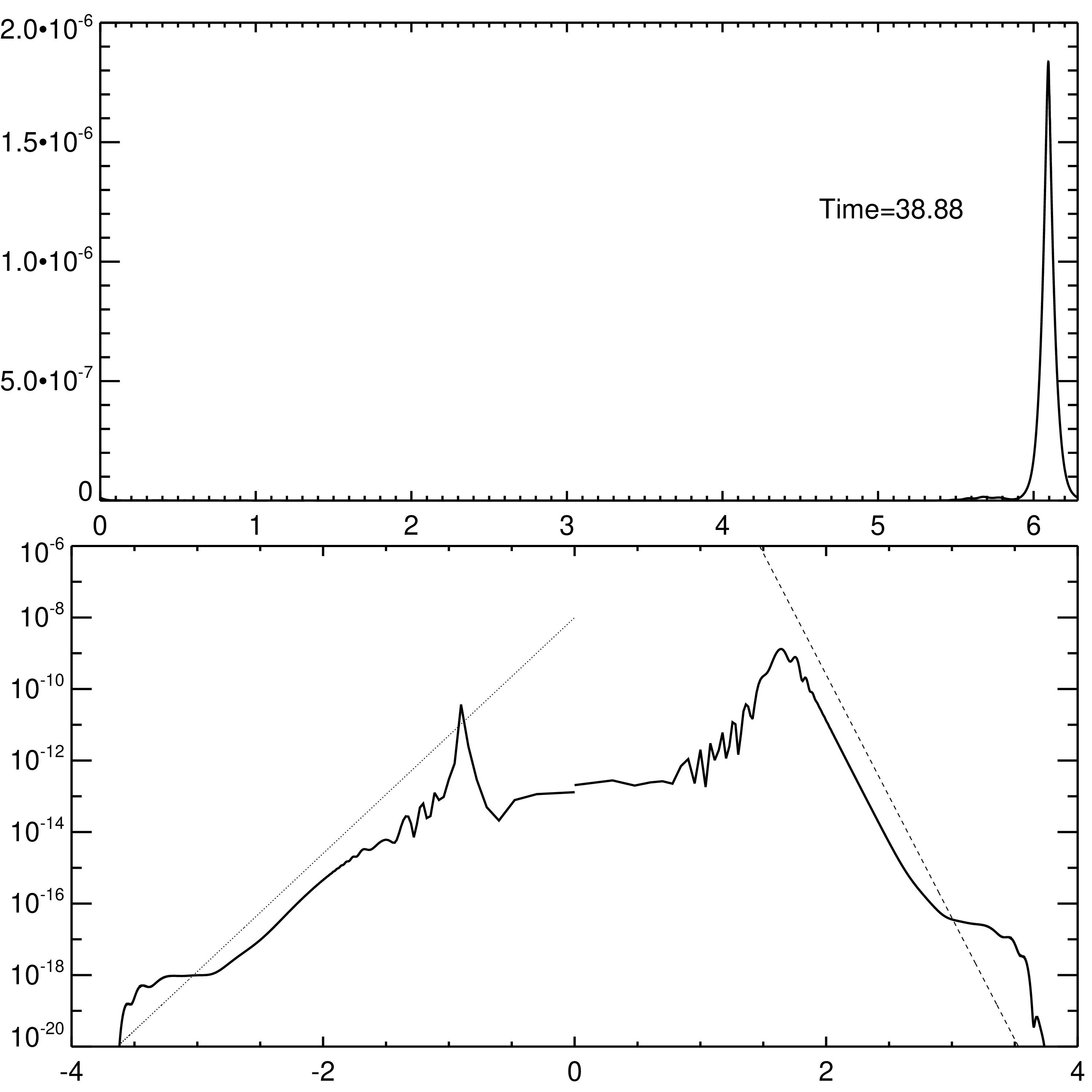} 
\caption{Same as Fig.\ref{InCond}, but for $t=38.88$, corresponding to the 1st maximum from Fig.\ref{OSC_MOM}. Left slope of the spectrum is approximated by function $\sim k^{-3.3}$ (dotted line), right slope is approximated by function $\sim k^{-6.8}$ (dashed line).} \label{Max}
\end{figure}

Fig.\ref{Max} presents real and Fourier space of the system at $t=38.88$, corresponding to the first maximum from Fig.\ref{OSC_MOM}. The real space picture of $|\psi(x)|^2$ shows that initial condition moved to the right with respect to initial condition, growing in the amplitude and narrowing in width. Also, small portion of the initial condition has been separated in the form of the hump of much smaller amplitude. Fourier space contains two maxima: the right major peak approximately at $k_m=50$, corresponding to the quasibreather, and the left smaller peak corresponding to the solution of the Eq.(\ref{C}):
\begin{eqnarray*}
 k_0 = -(3-\sqrt{8})\cdot k_m \simeq -8.6
\end{eqnarray*}
As shows Fig.\ref{Max}, the spectrum remains localized near initial wave number $k \simeq k_m$. This fact can be explained by conservation of both wave action and momentum. Thus, the turbulence inside the solution can be interpreted as an ``envelope turbulence''. It is interesting that the area of this turbulence is localized both in real and Fourier spaces.

Comparison with initial data shows that the spectrum  gains power-like tails $I_k\simeq k^{-3.3}$ for negative $k$ and $I_k\simeq k^{-6.8}$ for positive $k$. Recall that the simple collapse theory predicts $I_k\simeq k^{-5}$. Anyway, appearance of power-like tails indicates violation of smoothness of $\psi(x,t)$.

\begin{figure}[ht]
\includegraphics[scale=0.3]{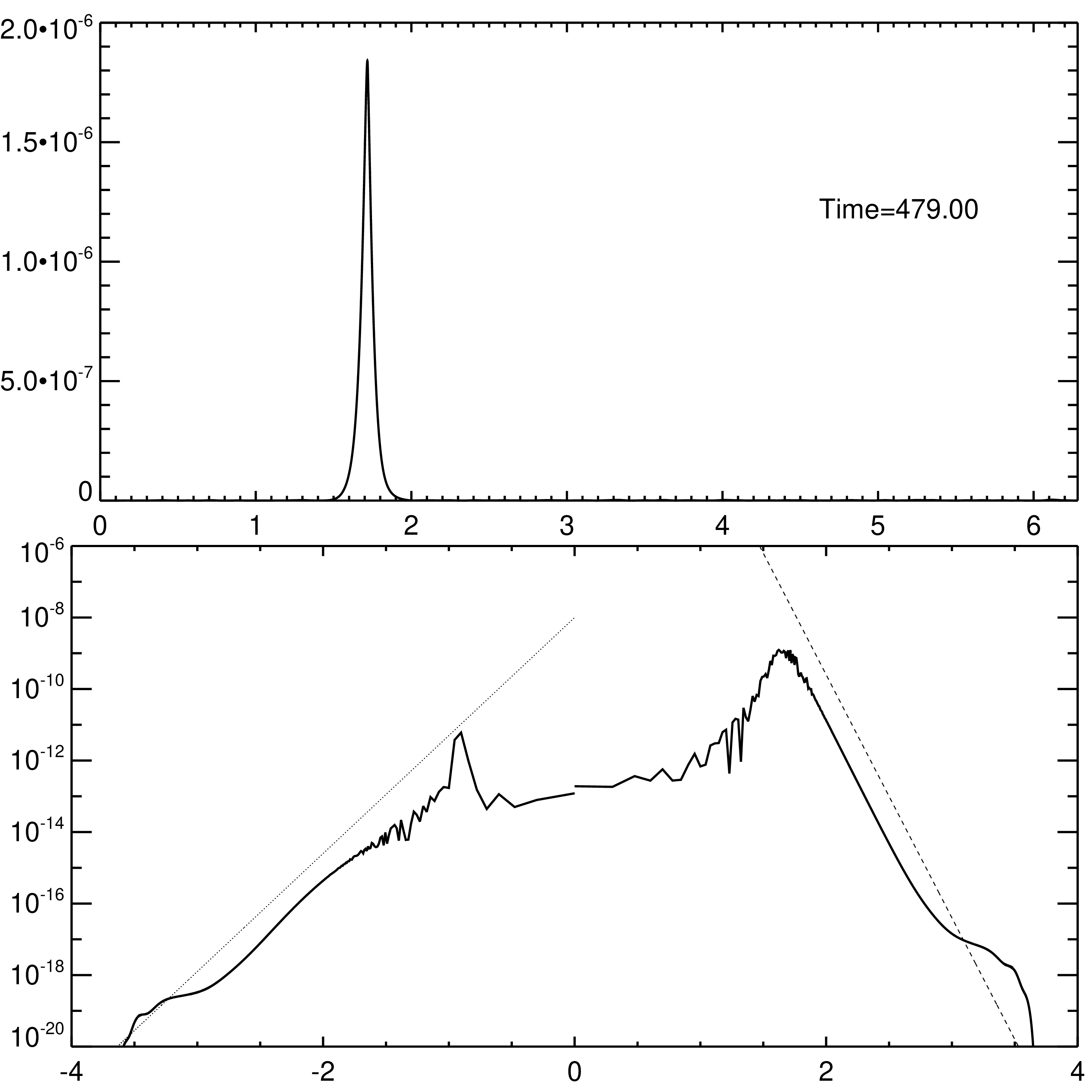} 
\caption{Same as Fig. \ref{InCond}, but for time $t=479.0$ corresponding to 6th maximum from Fig.\ref{OSC_MOM}. Left slope of the spectrum is approximated by function $\sim k^{-3.3}$ (dotted line), right slope is approximated by function $\sim k^{-6.8}$ (dashed line).} \label{6Max}
\end{figure}

The observed singularity is of weak-collapse type. It is confirmed by the fact that the amount of Hamiltonian absorbed during 11 periods of oscillations of quasibreather (see Fig.\ \ref{OSC_MOM}) is approximately equal to $0.03 \% $ of its initial value. One should note that observed picture is universal: another snapshots of the system, taken at the times corresponding to subsequent maxima from Fig.\ \ref{OSC_MOM}, reveal the pictures similar to observed on Fig.\ \ref{Max} (see, for example, Fig.\ \ref{6Max}). 
\noindent

\begin{figure}[ht]
\includegraphics[scale=0.3]{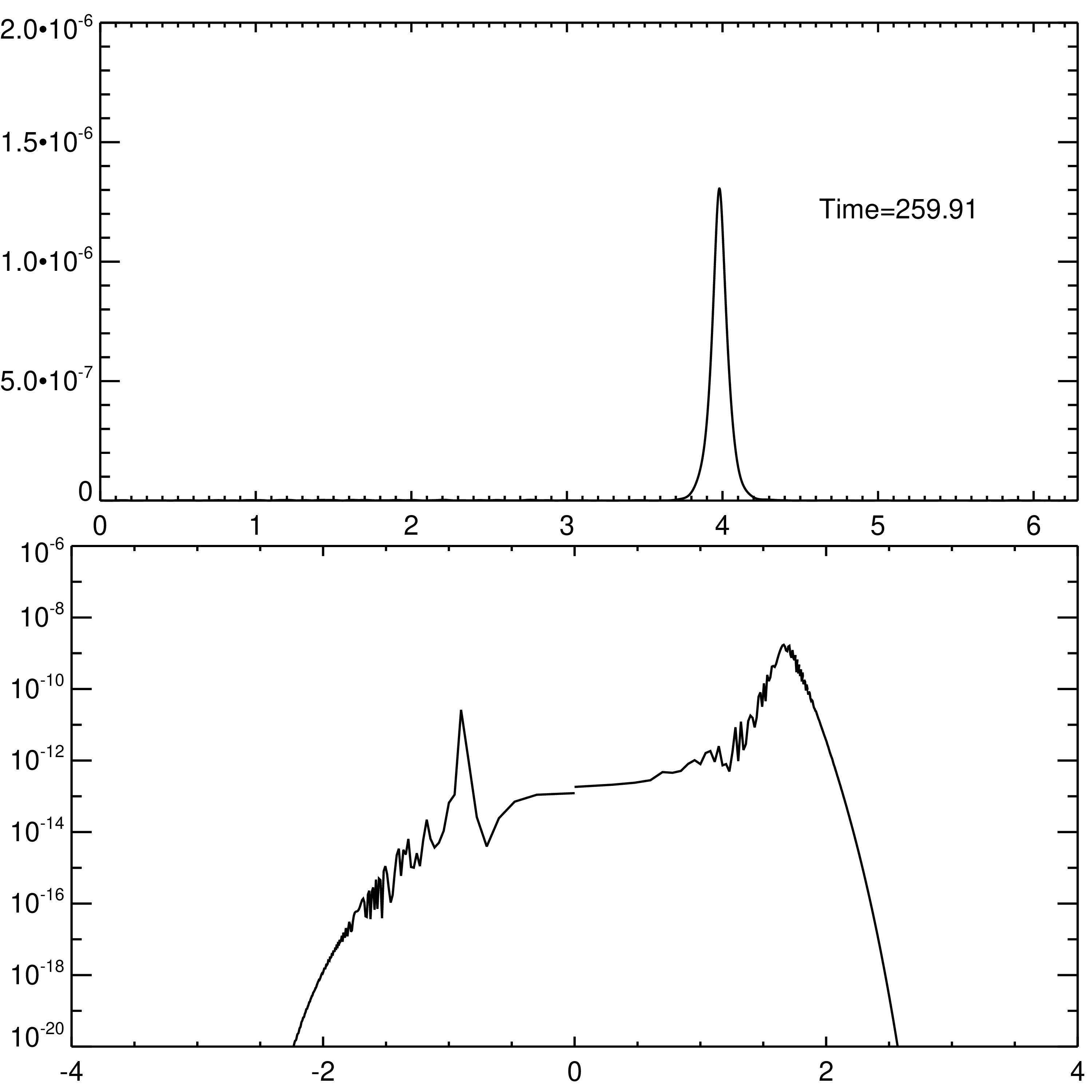} 
\caption{Same as Fig. \ref{InCond}, but for time $t=259.91$ corresponding to the 3rd trough from Fig.\ref{OSC_MOM}} \label{3Min}
\end{figure}
\noindent

Fig.\ref{3Min} presents real and Fourier space of the system at $t=259.91$, corresponding to the third trough from Fig.\ref{OSC_MOM}. The real-space picture of $|\psi(x)|^2$ shows that amplitude of quasibreather has been diminished with respect to the state corresponding to  Fig.\ref{Max}. Fourier space exhibits both similarities and differences being compared to bottom of the Fig.\ref{Max}: there are the same right main peak approximately at $k_m=50$ and the left smaller peak approximately at $k_0 = -8.6$, but high-wavenumber tails decay much faster than power law. It means that $\psi(x,t)$ is smooth at the moment of minimum.

\begin{figure}[ht]
\includegraphics[scale=0.4]{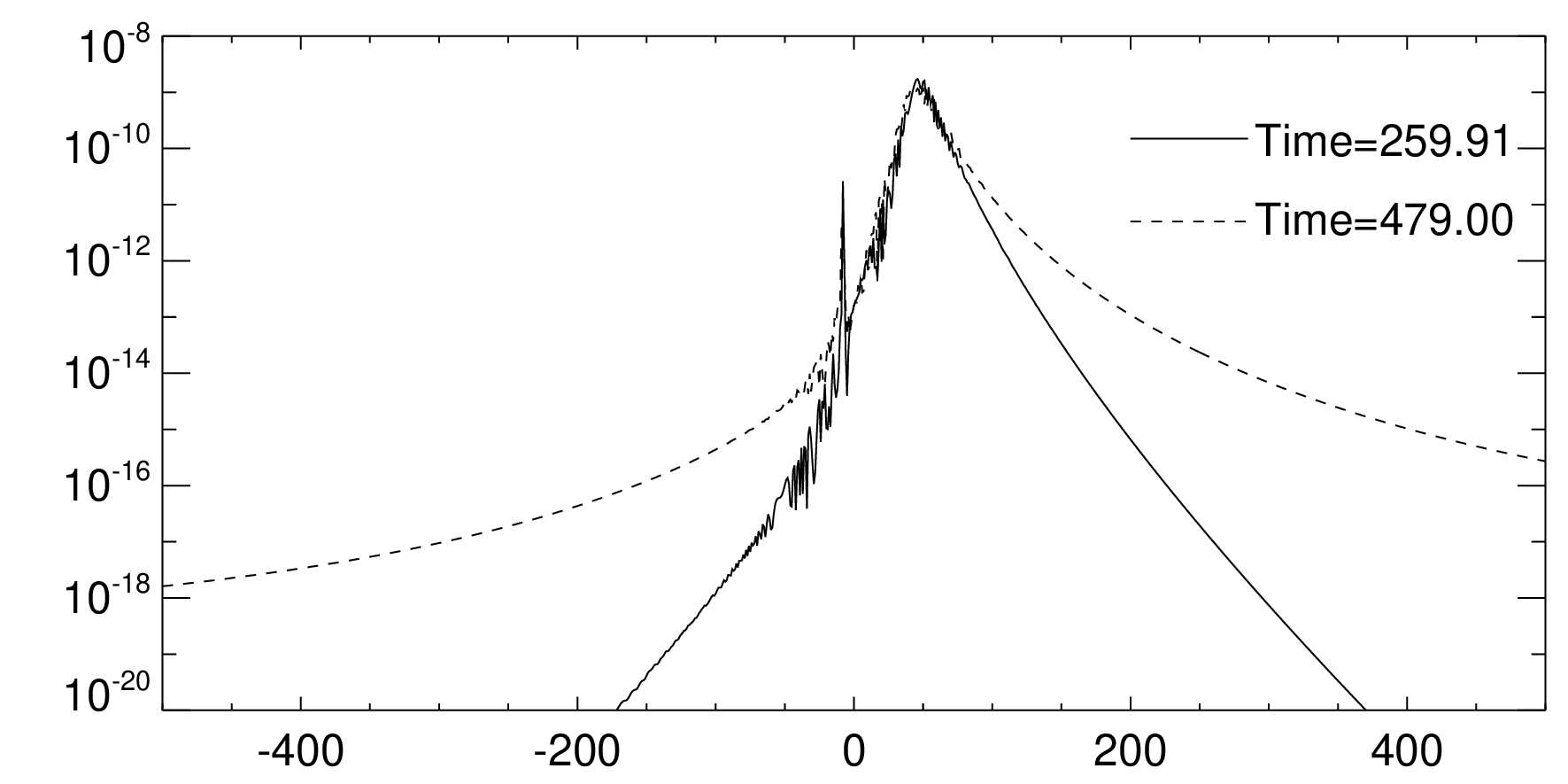} 
\caption{Comparison of two spectra $\log_{10} |\psi_k(t)|^2$ for time $t=259.91$ (solid line, corresponds to the third trough on the  Fig.\ref{OSC_MOM}) and time $t=479.00$ (dashed line, corresponds to the six' peak on the Fig.\ref{OSC_MOM}), plotted as a function of wave-number $k$. This picture demonstrates that the spectral tails "breath" between exponential and power-like states.} \label{Breathes}
\end{figure}

For the illustration of the quasibreather temporal behavior, we present Fig.\ \ref{Breathes}, showing two states of the system taken at the moments when quasibreather reaches it's maximum and minimum amplitude in semi-log scale. It's quite obvious that spectral tails decay exponentially at the moment corresponding to the amplitude minimum of quasibreather, and decay as a power of wave number at the moment of the quasibreather amplitude maximum. This solution, therefore, periodically "breathes" between states of singularity formation and its regularization.

\begin{figure}[ht]
\includegraphics[scale=0.4]{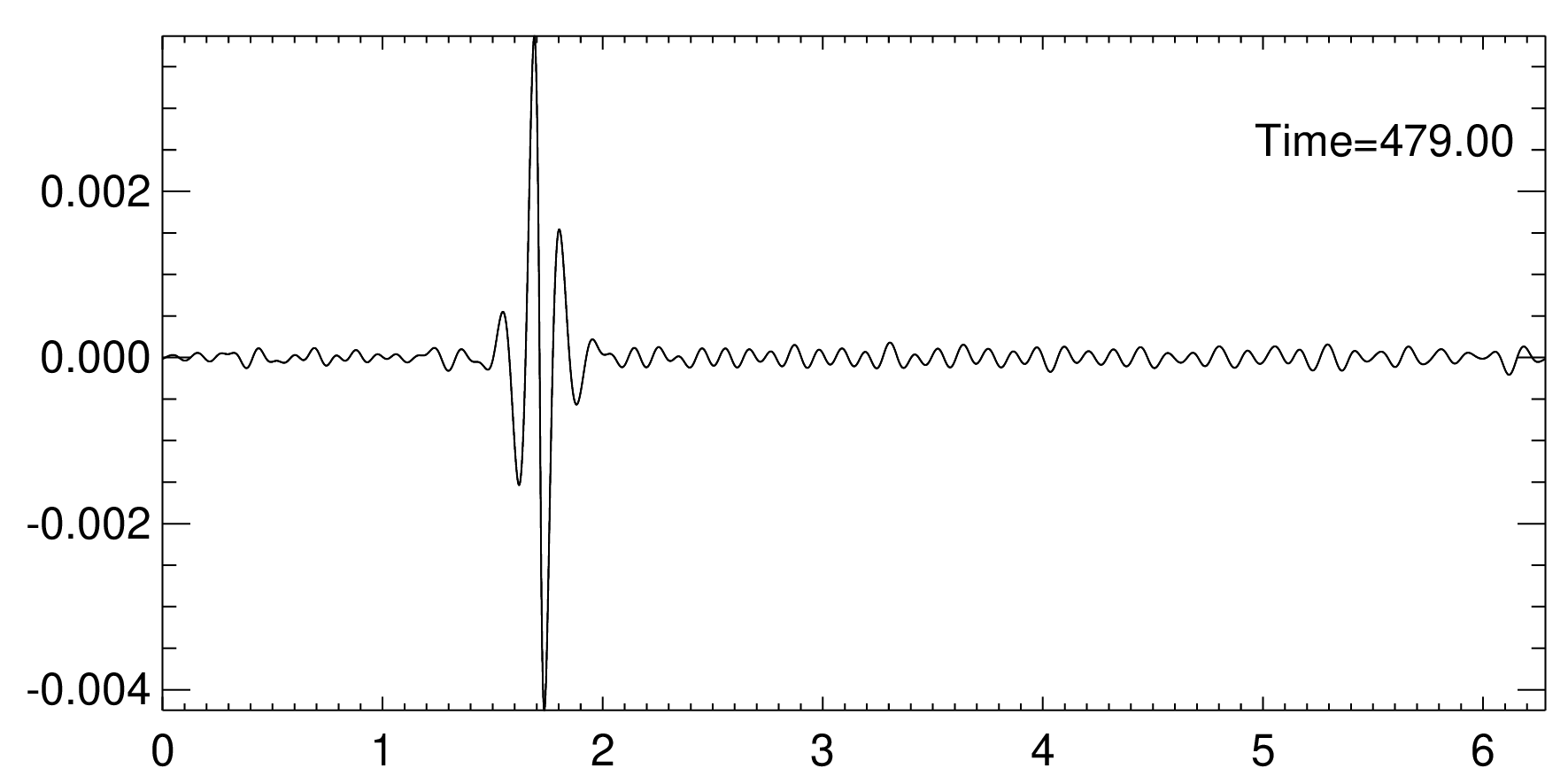} 
\caption{Surface elevation $\eta(x,t)$ as a function of real space coordinate $x$ for time $t=479.00$, corresponding to Fig.\ \ref{6Max}}\label{ThreeSisters}
\end{figure}

\begin{figure}[ht]
\includegraphics[scale=0.4]{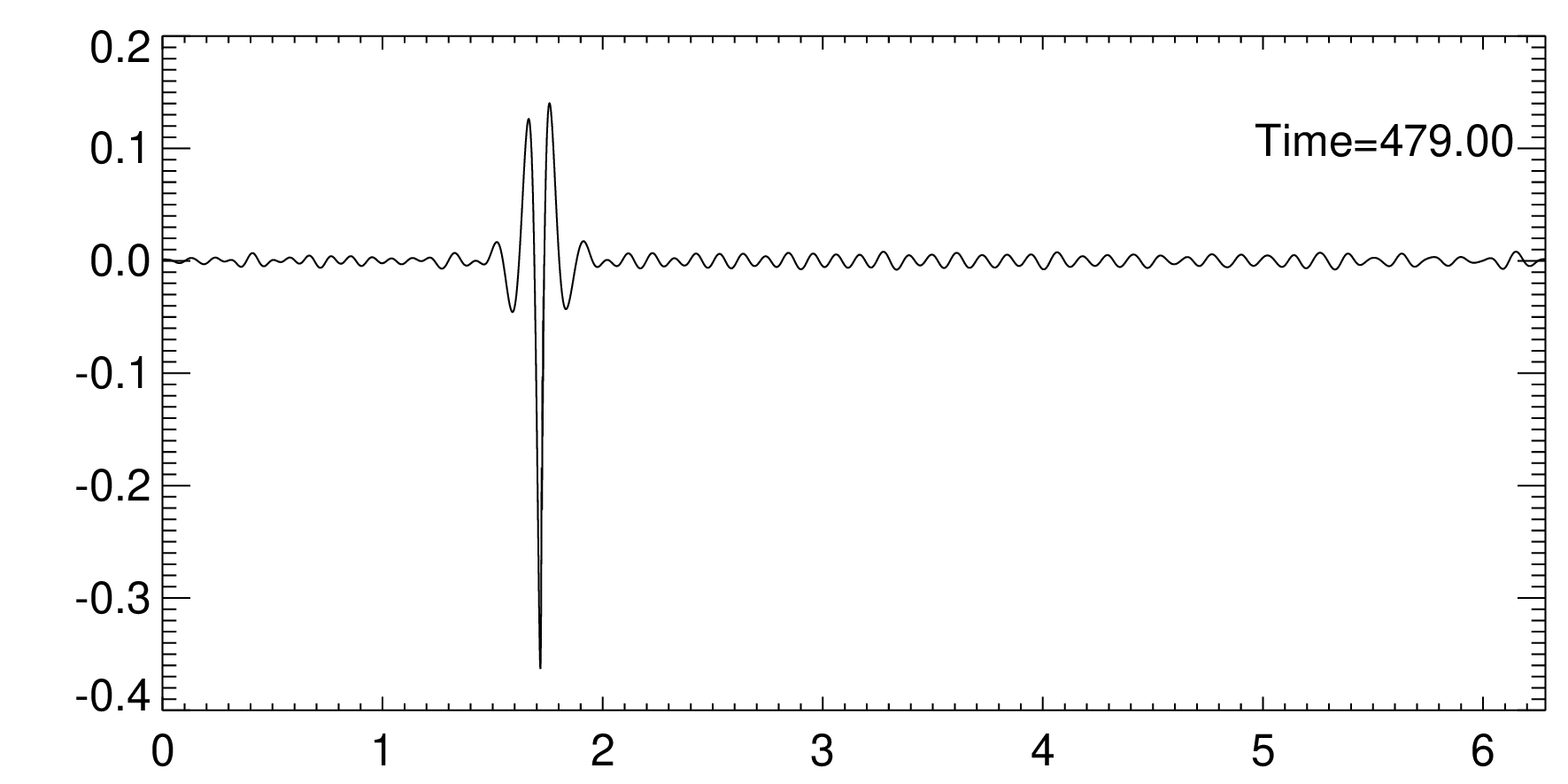} 
\caption{Slope of the surface elevation $\left.\frac{\partial \eta(x,t)}{\partial x}\right|_{t=479.0}$ as a function of real space coordinate $x$, corresponding to Fig.\ \ref{ThreeSisters}}\label{3_Sisters_Deri}
\end{figure}

Fig.\ \ref{ThreeSisters} presents surface elevation Eq.\ (\ref{SurfaceElevations}) for the same time as Fig.\ \ref{InCond}. This picture looks qualitatively similar to experimentally observed "Three Sisters" killer wave on the ocean surface \cite{PK} and the resent results on freakon simulation on the deep water surface \cite{ZD}. Fig.\ \ref{3_Sisters_Deri} shows slope elevations, corresponding to Fig.\ \ref{ThreeSisters}. These slope elevations values have the meaning of the original Euler equations for deep water surface gravity waves.

\noindent

\noindent

One remarkable feature of the observed quasibreather is its co-existence with surrounding noise environment, associated with the radiation at the secondary spectral peak at $k_0=-8.6$. In fact, the surrounding weakly-nonlinear noise could consists not only of radiation at wavenumber $k_0=-8.6$, but also of the products of the initial condition decay into quasibreather and other waves. However, the wave action density in this noise is so small with respect to energy density in quasibreather, that this noise certainly cannot be interpreted as a kind of "condensate".   

To analyze this situation, we performed the following experiment. In the middle of the simulation the real-space, containing quasibreather and surrounding noise, was "cleaned-up" through zeroing the function $\psi(x)$ everywhere except the carrier domain of the quasibreather. As a result, further evolution of the system starting from such "cleaned" initial conditions didn't show any qualitative difference from previous behavior -- we observed immediate appearance of the surrounding noise at $k_0=-8.6$ of the same characteristic amplitude, as we have seen before the "cleaning" of the real space.

This observation lead us to the conjecture that quasisolitons and quasibreathers exist only in quasi-equilibrium with weakly nonlinear wave noise environment.

Another important observation, which distinguishes quasibreathers from oscillations of perturbed $NLSE$ solitons, is periodical singularity formation at every time quasibreather reaches its maximum. This property is illustrated by both  Fig. \ref{Max} (corresponds to the first maximum from Fig.\ \ref{OSC_MOM}) and Fig. \ref{6Max} (corresponds to the maximum number six from Fig. \ref{OSC_MOM}).

In a nutshell, the gravity surface waves $MMT$ model shows periodic focusing of the initial condition Eq.(\ref{InCondSoliton}) with weak-collapse singularity formation exhibiting itself in power spectral tails and weakly nonlinear radiation at secondary spectral maximum at $k_0=-8.6$, which differs observed quasibreather from previously known breather-like structures. The similarity  of observed quasibreather in terms of water surface elevation with experimental "Three Sisters" wave packet and numerically observed freakon shows that even simplified model of gravity surface waves as $MMT$ catches significant properties of the original exact equations.

\section{Conclusion}
On the base of numerical experiments we see that quasisolitons in frame of defocusing $MMT$ model with parameters $\alpha=1/2$, $\beta=3$ and $\lambda=1$ are robust long-living objects, existing for hundreds of leading wave periods. Quasisolitons of large amplitude turn to quasibreathers. Their amplitude and spectral shape oscillate in time. These oscillations are accompanied by formation of weak collapses which can be compared with "white capping" of real ocean waves.

We conclude that the "solitonic" scenario of freak waves is based on the equal foot with alternative "instantonic" scenario. We need to perform more numerical experiments in the frame of exact Euler equation to establish what  scenario is closer to reality.

Let us mention that oscillatory effects in solitons propagating on zero background were observed in paper \cite{A3}. However, in this paper the authors studied not single $NLSE$, but the system of coupled $NLSE$. The dynamics of this system is much more complicated.

\section{Acknowledgments}
This work was sponsored by ONR grant N00014-10-1-0991, NSF grant \# 1130450, Russian Government contract $11.9.34.31.0035$, RFBR grant 12-01-00943, the Program of the RAS Presidium ``Fundamental Problems of Nonlinear Dynamics in Mathematical and Physical Sciences'' and the "Leading scientific schools of Russia" grant NSh 6170.2012.2. Authors gratefully acknowledge continuous support of these foundations.

\section{Appendix I}

Now we address the following question -- what value of $\lambda$ has to be chosen to provide the best possible modeling of real surface gravity waves on deep water?

To answer this question, we notice that weakly nonlinear gravity waves on deep water surface with gravity acceleration $g=1$  are described by so-called "Zakharov equation", which is exactly Eq.(\ref{MMT_FFT}) at $\alpha=1/2$.

The "real" coupling coefficient $T_{k k_1 k_2 k_3}$ is a complicated homogeneous function of the third order:
\begin{eqnarray}
T_{\epsilon k \epsilon k_1 \epsilon k_2 \epsilon k_3}^R=\epsilon^3 T_{k k_1 k_2 k_3}^R
\end{eqnarray}
Explicit expression for "real" $T_{k k_1 k_2 k_3}$ was found, for instance, in the paper \cite{PRZ}.

Functions $T_{k k_1 k_2 k_3}^R$ from \cite{PRZ} and $T_{k k_1 k_2 k_3}$, given by Eq.(\ref{MMT_ME}), are essentially different. However, we can make them coincide in one point $k=k_1=k_2=k_3$ by the proper choice of $\lambda$.

According to \cite{PRZ}
\begin{eqnarray}
\label{MC}
T_{k k_1 k_2 k_3}^R= \frac{1}{4 \pi^2} k^3
\end{eqnarray}
But in the cited paper we used the "symmetric form" of the Fourier transform. If we define the Fourier transform according to Eq.(\ref{FT}), we must replace Eq.(\ref{MC}) to 
\begin{eqnarray}
T_{k k k k}^R = k^3
\end{eqnarray}
Hence, to reach the best approximation to reality, we have to put
\begin{eqnarray}
T_{k k k k} = k^3
\end{eqnarray}
It means that we must choose $\lambda=1$. Then the shape of the surface $\eta(x,t)$ defined by Eq.(\ref{SurfaceElevations}) is a model (rather approximate, of course) of a real water surface. From Fig. \ref{ThreeSisters} one can conclude that the steepness of our breather is fairly high and hardly can be described by $NLSE$.

\section{Appendix II}

The vast majority of surface waves physical characteristics measurements is coming from stationary installations like oil platforms, presenting the water surface elevations as a time series.

The water surface elevation itself is not a measure of the system nonlinearity degree, since underlying equations are invariant with respect to stretching transformations, therefore surface waves of height varying by the order of magnitude can be of the same degree of nonlinearity. 

The real physical characteristic of nonlinearity is the wave slope $\mu$, which needs to be recovered from the surface elevations time series. Here we preset such simple estimate.

By definition, the slope (same as steepness) is $\mu = ka$, where $k$ and $a$ are the characteristic wave number and amplitude correspondingly. The connection between wave period $T$ and wave number is 
\begin{eqnarray}
k=\frac{4 \pi^2}{g T^2}
\end{eqnarray}
For the famous "Draupner Wave" (also known as "New Year Wave", see Fig.\ref{Draupner}, \cite{DW}), $T=12\,sec$, $a=13.7\, m$ and $g=9.81 \, m/sec$, which gives $\mu \simeq 0.38$ in accordance with our experiments.

\end{document}